\documentstyle[12pt]{article}
\newcommand{\be}{\begin{equation}}
\newcommand{\ee}{\end{equation}}
\newcommand{\bea}{\begin{eqnarray}}
\newcommand{\nn}{\nonumber}
\newcommand{\eea}{\end{eqnarray}}

\begin{document}
 
\begin{titlepage}
\begin{flushright}

UA-NPPS-00-15 \\
hep-th/0007207 \\
\end{flushright}

\begin{centering}

\vspace{.1in}

{\large {\bf Static and non-static quantum
effects in two-dimensional
dilaton gravity }} \\

\vspace{.2in}

 {\bf C. Chiou-Lahanas}, {\bf G.A. Diamandis}, {\bf B.C. Georgalas},
 {\bf A. Kapella-Ekonomou} 
and {\bf X.N. Maintas} \\
  {\it University of Athens, Physics Department, Nuclear and Particle
  Physics Section, Panepistimioupolis, Ilisia 157-71, Athens, Greece}

%


\vspace{.5in}
 
{\bf Abstract} \\
\vspace{.1in}
\end{centering}
{\small We study backreaction effects in two-dimensional dilaton gravity.
The backreaction comes from an $R^2$ term which is a part of the
one-loop effective action arising from massive scalar field quantization in
a certain approximation. The peculiarity of this term is that it
does not contribute to the Hawking radiation of the classical
black hole solution of the field equations. In the static case we
examine the horizon and the physical singularity of the new black hole
solutions. Studying the possibility of time dependence we see the
generation of a new singularity. The particular solution found still
has the structure of a black hole, indicating that non-thermal effects
cannot lead, at least in this approximation, to black hole evaporation. }

\end{titlepage}

\section{Introduction} 

 Two-dimensional dilaton gravity is a useful laboratory
 for addressing fundamental questions of quantum gravity.
 It exhibits many of the interesting non-trivial features of 
 four dimensional gravity, describing the s-wave sector of
 four dimensional Einstein gravity, and especially black hole
 solutions \cite{callan} permiting a study of
 the interplay between gravity
 and quantum mechanics. This fact combined with the stringy
 origin of the action justifies the persisting activity in the
 field.

 \par
 In a previous work \cite{hawking} the Hawking radiation due to the
    presence of
 quantum scalar massive particles has been calculated. One of 
 the results of this work was that purely geometrical terms
 (not including the dilaton field) in the one-loop effective
 action do not contribute to the thermal radiation. This
 observation raises the question of the evolution of a 
 Schwarzschild black hole in the presence of such terms.
 The first of these terms in the one-loop effective action,
 in a large mass expansion, is just an $R^2$ term.
 Thus two-dimensional higher derivative gravity is interesting
 for one more reason. It can in principle give information
 about non-thermal effects of quanrum origin, in black hole pfysics.
 Backreaction
 effects including this term were considered in \cite{backre},
 where static black hole geometries were found as solutions
 of the field equations of the two dimensional dilaton gravity
 with $R^2$ term. The existence of these solutions viewed as
 backreaction effects is compatible with the fact that the extra
 terms do not contribute to the Hawking radiation but in no way
 can say something about the evolution of the original black hole
 like the case, for example, of the Russo, Susskind, Thorlacius
  model \cite{rst}.
 \par
 In this work we try to incorporate time dependence in the 
 backreaction effects due to the presence of the $R^2$ term.
 In section 2 we formulate the problem and give the field
 equations both in covariant form and in the conformal gauge
 using light cone coordinates, which are more convenient to
 incorporate time dependence. In section 3 we point out
 the basic features of the static solutions found in
 \cite{backre}. The analysis here is presented in the
 abovementioned coordinate system giving
 emphasis in the study of the region behind the event
 horizon and in the appearence of the singularity. This discussion
 completes the results of \cite{backre}. In section 4 we
 consider the non-static case. We find time-dependent solutions
 expanding in powers of the coefficient ($\kappa$) of the $R^2$ term .
 In $\kappa^2$ order we find solutions with black hole structure
 differing from the static case mainly in the appearence of
 an extra physical singularity. Up to this order the solution found
 has striking similarity to the one found in \cite{exact}.
 In this reference the corresponding solution was an
 exact solution of the field equations in the presence of
 non-trivial (classical) tachyon configuration. We conclude
 with a discussion of our results in section 5.

 \section{Setting the problem}
 
 The classical action of a scalar field coupled to the two-dimensional
 gravity (and to the dilaton field) is
 
 \be
 S_{cl} = \frac{1}{2\pi}\int \sqrt{-g} e^{-2\phi} \{ [R+4(\nabla \phi)^2
 + 4\lambda^2] - [(\nabla T)^2 - m_0 T^2] \}
 \label{class}
 \ee

 while the rescaling $T = e^{-\phi} \tilde{T}$ giving to the scalar field
 canonical kinetic terms yields:

 \be
 S_{cl} = \frac{1}{2\pi}\int \sqrt{-g} e^{-2\phi} \{ [R+4(\nabla \phi)^2
 + 4\lambda^2] -  
 [ (\nabla \tilde{T})^2 + ( (\nabla \phi)^2 - \Box \phi -
 m_0 ^2 ) \tilde{T}^2 ] \}.
 \label{scaled}
 \ee

 Note that the term $ ( (\nabla \phi)^2 - \Box \phi - \lambda^2)
 \tilde{T}^2 $
 gives the quadratic coupling of the scalar field to the dilaton, being zero
 in the linear dilaton vacuum,  while
 $ m^2 = \lambda^2 - m_0^2 $ is the "mass" of the scalar field.
 
 \par
 Quantizing the scalar field one can take the one-loop
 effective action, which contains both local and non-local
 terms \cite{avramidi}, \cite{hawking}. 
 Expanding the purely geometric part of the non-local terms
 (which are responsible for the Hawking radiation)
 in inverse powers of the mass $m^2$ and keeping the first term 
 \cite{backre} we get

 \be
 S = S_{cl} - \kappa \int \sqrt{-g} R^2
 \label{action}
 \ee
where $\kappa=1/(240m^2)$. Since we are not interested in this work in
non-trivial
 scalar field configurations we consider the metric and dilaton equations  
 which read:

 \be
 -2 e^{-2\phi} [g_{\mu \nu} ( (\nabla \phi)^2 - \Box \phi - 1 ) +
 \nabla _\mu \nabla _\nu \phi ] - 
 \kappa [ 2 \nabla _\mu \nabla _\nu - \frac{g_{\mu \nu}}{2} R^2 - 2
 g_{\mu \nu} \Box R] = 0,
 \label{covmetric}
 \ee

 \be
 e^{-2 \phi} \{ \frac{R}{4} - [(\nabla \phi)^2 - \Box \phi - 1 ] \} =0
 \label{covdilaton}
 \ee

 In the following we'll work in the conformal gauge and light-cone
 coordinates $(x^+,x^-)$ where the line element reads
 \be
 ds^2=-e^{2\rho}dx^+dx^-
 \ee
and the components of the Ricci tensor take the form
 $R_{++} = -2\partial_+ ^2 \rho$, $R_{--}=-2\partial_- ^2 \rho$ and
 $R=8e^{-2\rho}
 \partial _+ \partial_-\rho$ while for a scalar field $ f $ we have
 $\Box f = -4 e^{-2 \rho} \partial _+ \partial_- f $.

 In this gauge the dilaton equation becomes
 
 \be
 e^{-2 \phi} \{ 1+ 2e^{-2\rho}\partial_+\partial_-\rho
 + 4 e^{-2 \rho}[\partial_+
 \phi \partial_- \phi
 - \partial_+\partial_- \phi] \} = 0
 \label{dilaton}
 \ee
while from the equations for the metric components we get the trace equation
 
 \be
 -e^{2(\rho - \phi)} + e^{-2 \phi} (-4 \partial_+ \phi \partial_- \phi
 +2 \partial_+\partial_- \phi) + \kappa (2 \partial_+\partial_ -R -
 \frac{1}{4} e^{2\rho} R^2) = 0
 \label{trace}
 \ee
 and the two constraints

 \bea
 e^{-2\phi}(2 \partial_-^2 \phi - 4 \partial _-\phi \partial _-\rho) + \kappa
 (4 \partial _- \rho \partial _- R - 2 \partial _- ^2 R) &=& 0  \\
  e^{-2\phi}(2 \partial_+^2 \phi - 4 \partial _+\phi \partial _+\rho) + \kappa
 (4 \partial _ +\rho \partial _+ R - 2 \partial _+ ^2 R) &=& 0 
 \label{constraints}
 \eea

 Note that combining the trace (\ref{trace})
 and the dilaton (\ref{dilaton}) equations we get  the
 relation

 \be
 2 e^{-2 \phi} \partial_+ \partial_- (\rho - \phi) = \kappa
 [ 16 e^{-2 \rho} ( \partial_+ \partial_- \rho )^2 - 
 16 \partial_+ \partial_- (
 e^{-2 \rho} \partial_+ \partial_- \rho)].
 \ee

 We see that when $\kappa=0$ we recover the $\rho - \phi$
 symmetry \cite{callan},\cite{rst}.
 The system
 admits in this case  the solution

 \be
 e^{-2 \phi} = e^{-2 \rho} = M - x^+x^-
 \label{staticc}
 \ee
for the conformal factor and the dilaton field while for the Ricci
 scalar we take:

 \be
 R_{\kappa=0}=4Me^{2 \rho}.
 \ee 
 The solution in (\ref{staticc}), unique in the $\kappa=0$ case,
 describes either a static black hole solution ($M \neq 0$) or flat
 space
 ($M=0$) \cite{callan}. The presence of the backreaction terms
 ($\kappa \neq 0$), spoils the
  $\rho - \phi$ symmetry leading to new static solutions and permitting
  time dependence.

 \section{Static Solutions}

 Static black hole solutions emerging from the Lagrangian in
 (\ref{action}) have already been found in \cite{backre} (and
 in somehow different context in \cite{odintsov}) where
 the Schwarzschild coordinate system was adopted. In this section
 we will reconsider the basic features of these 
 solutions working in the conformal gauge mainly for use in the
 time dependent case. 

 For static solutions we consider the fields as functions of the
 combination $-x^+x^-$ only. This is clear from
 the fact that the coordinates $x^{\pm}$  are
 related to  the  physical spacetime conformal
 coordinates through the relations  $x^+=e^{\sigma ^+}$,
  $x^-=-e^{\sigma ^-}$,
 where $\sigma^{\pm}= \frac{1}{2}(t \pm x)$, or  conversely $x=ln(-x^+x^-)$, 
 $t=-ln(-\frac{x^-}{x^+})$.
 
 The system of  the equations is a fourth order one, due to the presence
 of the $R^2$ term in the action. In order to reduce the order we assume
 the field
 $\rho$ is monotonic function of the variable $x^+x^-$. This
 is always possible since the number of degrees of freedom
 ($\rho$ and $\phi$ in our case) is less than the number of
 equations so one of the degrees of freedom can have this
 property. Note that the dilaton field cannot be a monotonic function
 as we have seen from the analysis in \cite{backre}.
 The above monotonicity property is also true in four-dimensional
 black hole solutions
 \cite{canti}. The non-monotonicity of the dilaton
 is explained by the fact that the $R^2$ forces are
 repulsive. Furthermore we introduce
 the "auxiliary" field $A(\rho)$ through the relation:
 
 \be
 R(x^+, x^-) = 8e^{-2 \rho} \partial_+ \partial_- \rho = 8 e^{-2 \rho} A(\rho).
 \label{auxcurv}
 \ee
 
 and we also consider  the dilaton field being function of $\rho$,
 $\phi(\rho)$.
 Writting the system of equations in terms of $A(\rho)$ and $\phi(\rho)$ and 
 eliminating the derivatives of $\rho$ we are left with an ordinary non-linear
 system. For this system we seek solutions analytic in the parameter
 $\kappa$
 written in the form of $\kappa$-series as
 
 \bea
 A(\rho) & = & A_1(\rho) + \kappa A_2(\rho) + ...  \nonumber \\
 \phi (\rho) & = & \rho + \kappa H_1 (\rho) + \kappa^2 H_2 (\rho) + ...
 \label{staticans}
 \eea
where $H(\rho)=\kappa H_1 (\rho) + \kappa^2 H_2 (\rho) + ...$ denotes
 the deviation from the $\rho - \phi$ symmetry in the presence of the
 $R^2$ terms. For convenience in the integration of the equations we use 
 one more auxiliary  field:
 
 \be
 h(\rho)=e^{-2 \rho} - F(\rho ; \kappa) = M - x^+x^-.
 \label{auxrho}
 \ee

 The solution of the resultant system up to $ \kappa^4 $ -order reads:
 \bea
 A(\rho) & = & M_{10} e^{4 \rho} [ \frac{1}{2}+16 \kappa e^{2\rho} 
                +  16 \kappa^2 (96-M_{10}) e^{4\rho} +
           \frac{64(41472-733M_{10})}{9} \kappa^3 e^{6\rho} ]  \nonumber \\
 \phi(\rho) & = & \rho + \kappa M_{10} e^{4\rho} 
        [1  +   \frac{512}{9} \kappa e^{2\rho}
             + \kappa^2 \frac{41472-437M_{10}}{6}e^{4\rho} ]
  \label{kappasol}
 \eea
while the dependence of the fields on the spacetime coordinates is
 given through the field $h(\rho)$ up to $\kappa^2$-order by

 \be
 h(\rho) = e^{-2\rho} -8\kappa^2 M_{10} e^{2\rho} 
 + O(\kappa^3) = M - x^+x^-
 \ee

 The expression of the solutions in $\kappa$-expansion will be proven
 helpful in the time-dependent case but since it appears arbitrary
 \footnote{In fact analitycity in $\kappa$ comes from the general
 theory for non-linear ordinary differential systems.}
 we will justify it before proceeding in the discussion of the solutions.
 Since the solutions we seek for have to meet the asymptotic flatness
 condition, the fields $A(\rho)$, $H(\rho)$ have vanishing limit
 as $ \rho \rightarrow - \infty $. So we can solve the nonlinear system
 of equations by iteration following the method adopted in \cite{backre}. 
 The linearized system reads:

 \bea
 H^{'''} & = & 2e^{-2\rho}(A^{'} -4A)
 +8\kappa(-A^{''}+4A^{'}-4A) + H^{''}   \nonumber \\
 A^{''} & = & 4A^{'}-4A + \frac{1}{8\kappa}H^{''}
 \label{alpha}
 \eea
which after the elimination of the field $H$ yields an equation for $A$
 namely
 
 \be
 (4A-A^{'})+4\kappa e^{2\rho}(4A^{'}-4A^{''}+A{'''}) = 0.
 \ee
 
 The solution of this equation is:

 \be
 A(z)=\frac{c_1}{z} I_1(z) + \frac{c_2}{z} K_1(z) + 
 c_3Im [\frac{1}{z} S_{-2,1}(iz)]
 \label{alpha1}
 \ee
where $z=\frac{e^{-\rho}}{2\sqrt{\kappa}}$, $I_1$, $K_1$ are the modified
 Bessel functions, $S_{-2,1}$ is the Lommel function and the $c_i$'s are
 integration constants.
 \par
 Now if we write the system as a quasi-linear first order system

 \be
 \vec{X}'(\rho) = {\bf B(\rho)} \vec{X} (\rho) +
 \vec{{\bf F}}(\vec{X}),
 \label{newsys}
 \ee
 where ${\bf B}$ is a $5\times 5$ matrix giving the linear part of the
 system,

 \be
 {\bf \vec{F}}^{\mathrm{T}} = (0,0, {\bf F}^H, 0, {\bf F}^A)
 \ee
denotes the non-linear terms, and

 \be
 \vec{X}(\rho)^{\mathrm{T}}
 =(H(\rho),H^{'}(\rho),H^{''}(\rho), A(\rho), A^{'}(\rho)).
 \ee

 The solution in (\ref{alpha1}) can be used to produce a
 fundamental solution ${\bf Y(\rho)}$ of the linear part of (\ref{newsys}).
Then the
 full (asymptotically stable)
 solution is given iteratively in the form:

 \be
 \vec{X}_n(\rho)=\vec{X}_{(n-1)}(\rho)+
  \int ^{\rho} {\bf Y}(\rho) {\bf Y}^{-1}(t)
   {\bf \vec{F}}[\vec{X}_{(n-1)}(t)]dt
 \ee
 We don't present here the corresponding
 expressions because they are too lengthy.
 Working out the first iteration we see that the $\kappa$-series solution 
 (\ref{kappasol}) coincides with the asymptotic expansion
 of the exact solution
 taken by the iteration scheme confirming, at least asymtotically,
  the analyticity in the parameter $\kappa$.
 \par
 Note that the position of the apparent horizon (coinciding with the event
 horizon in the static case) is given by
 \be 
 \partial_+ \rho = -x^- J(\rho) = 0,
 \label{horizon}
 \ee 
 where $J(\rho)=\frac{1}{\partial _{\rho} h(\rho)}$.
 In the case of the C.G.H.S. black hole \cite{callan} we have
 $J_c(\rho)=-\frac{1}{2}e^{2 \rho}$
 leading to
 the unique apparent horizon, coinciding with the event horizon, at $x_-=0$.
 In our case the function $J(\rho)$ is more complicated leaving room
 for other
 solutions of (\ref{horizon}). Substituting the derivatives of $\rho$ from
 the constaints in (\ref{constraints}) we  get the equation for the field
 $J(\rho)$ which keeping terms up to second order reads

 \be
 J^{'}(\rho)=2J(\rho) - 2J(\rho)H^{''}(\rho).
 \label{rho}
 \ee

 From the second of equations in (\ref{alpha}) and taking into account that
 $R^{(2)}(\rho)=8e^{-2\rho}A(\rho)$ the equation (\ref{alpha}) can be 
 written as

 \be
 J^{'}(\rho)=2J(\rho) - 2\kappa J(\rho) e^{2\rho}R^{(2)''}(\rho).
 \label{jequation}
 \ee

 The solution of the linear part is taken to be
 $J_{lin}(\rho)=J_{c}(\rho)=-\frac{1}{2}e^{2\rho}$.
 Since the field $J$ does not appear in the system (\ref{alpha}) it is
 easy to enlarge the system including also the
 equation (\ref{jequation}) and to work out the first iteration.
 Thus we can take an expression for the field $J(\rho)$. We see
 now that $J(\rho)$ has a zero at some   $x^+x^-=constant<0$.
 The new apparent horizon appears when $J(\rho_H)=0$.
 So we see that the backreaction effects shift the horizon  in the
 physical spacetime region of the C.G.H.S. case namely on an hyperbola 
 $x^+x^-=constant<0$.
 \par
 In order to investigate the existence of the singularity we will
 make some comments for the function $h(\rho)$ (in fact the
 implicit dependence of $\rho$ on the space coordinate).
 Using the ansatz in (\ref{staticans}) in the form $\phi(\rho)=
 \rho + H(\rho)$ the dilaton equation is written as:

 \be
  1+ \partial_+\partial_- e^{-2 \rho} +
  e^H \partial_+ \partial_- \{ \int^{\rho}
  e^{-2y-H(y)}[-4H'(y)]dy \} = 0.
  \label{dilh}
  \ee
Multiplying by appropriate factor the above equation can be
 formally integrated to give:

 \be
 h(\rho) = e^{-2\rho} + \int^{\rho} e^{-2y-F(y)}[-4H'(y)]dy 
 \ee
where $F(\rho)$ satisfies the equation

 \be
 e^{-F-H}[-F'(H'+2H'^2)+ (2H'^3+H'')]+(e^{-H})''=0.
 \label{equationf}
 \ee

 This equation can be solved exactly to give the function
 $F(\rho)$  in terms of $H(\rho)$. Instead of giving the
 lengthy expression for $F$ we investigate its behaviour
 near the singularity ($\rho\rightarrow\infty$). The relevant
 terms in (\ref{equationf}) give the (asymptotic) equation

 \be
 e^{-F} F'(-2H'^2)+e^{-F}(2 H'^3) = 0
 \ee
with solution

 \be
 F_{sing}(\rho) \sim H_{sing}(\rho).
 \ee
So the function $h(\rho)$ behaves near the singularity as:

 \be
 h_{sing}(\rho)=e^{-2\rho} + 4 e^{-2\rho-H} -2\int^{\rho}
 e^{-2y-H(y)}dy.
 \ee

  Since the function $e^{-H(\rho)}$ vanishes very rapidly
  near the singularity ($H(\rho)
  \rightarrow +\infty$ as $\rho \rightarrow +\infty$),
  we see that $h(\rho)$ remains a positive function going to zero
  as $\rho \rightarrow +\infty$. From the definition
  of $h(\rho)$ in (\ref{auxrho}) we see that this happens
  when $x^+x^-=M$.
  \newline
  As far as the asymptotically flat region is concerned
  ($\rho \rightarrow -\infty$) the equation (\ref{equationf})
  is approximated by:

  \be
  e^{-F}(H''-F'H')-H''=0
  \ee
admitting the general solution

  \be
  F_{as}(\rho)=-Log \left[ \frac{H'(\rho)-c}{H'(\rho)} \right]
  \ee
which with the choice $c=0$ gives
  \be
  F_{as}(\rho)=0.
  \ee

  Then as is easily recognized from the dilaton equation, keeping only
  linear terms in the field $H(\rho)$,

  \be
  h_{as}(\rho)=e^{-2\rho} + 2 \kappa e^{-2\rho}
  \int^{\rho} e^{2y}R''(y)dy - 2\kappa R'(\rho)
  \ee
  where we have used the fact that asymptotically
 \be
 H''(\rho)=\kappa e^{2\rho}R''(\rho).
 \ee

 Substituting from (\ref{alpha1}) the expression $R \sim z
 Im[S_{-2,1}(iz)]$ we find that

 \be
 h(z)=4\kappa \left[ z^2-\frac{z^2}{2} \int^z
 \frac{Re[S_{-1,0}(iu)]}{u}du \right].
 \ee

  The above analysis convinces about the monotonicity
  of $\rho$ as a function of $x^+x^-$ which was anticipated
  in the begining of this section. For the singularity we have
  to make two comments. The first is that in this analysis
  comes out naturally that the geometry of the spacetime
  has a physical singularity. This could not be seen
  from the numerical solutions in the Schwarzschild gauge
  as e.g. in
  \cite{perry}, \cite{backre}. The second remark is that
  unlike the C.G.H.S. case the constant $M$ appearing in
  (\ref{auxrho}) is not the mass of the black hole. Thus it
  can take negative values also. This means that there exist
  $R^2$ black hole solutions which have both their horizon
  and their singularity lying in different hyperbolae in the
  physical spacetime region of the C.G.H.S. geometry.

 \section{Time dependence}
 
 In order to incorporate the time dependence we keep $\rho$ as
 one of the variables but we allow explicit
 dependence on $x^+$ also.
 In particular the implicit dependence of $\rho$ on the
 variable $x^+x^-$ of the static case (\ref{auxrho}) 
 is replaced now by the ansatz:
 
 \be
 h(x^+, \rho)=G(x^+) - x^- F(x^+)
 \label{timeh}
 \ee
where $F(x^+)$ must be a monotonic function and the left hand side
 has the form:
 
 \be
 h(x^+, \rho) = e^{-2\rho} + \kappa h_1 (x^+, \rho).
 \label{timeh1}
 \ee
 
 When $\kappa = 0$  we get the well known solution with $G(x^+)$ 
 constant and $F(x^+) = x^+$.
 Having in mind an evolution scheme we make the following ansatz
 for the fields $\phi$ and $\partial_+\partial_-\rho$:
 
 \bea
 \phi &=& \rho + f_{st}(\rho;\kappa) + \kappa H(x^+, \rho) \\
 \partial_+ \partial_- \rho &=& A_{st}(\rho;\kappa) + \kappa
  A(x^+, \rho)
 \label{ansatz}
 \eea                                        
where $f_{st}(\rho ; \kappa)$ and $A_{st}(\rho; \kappa)$ are the
solutions found in the static case.
 \newline
 Now the system of equations (\ref{dilaton}, \ref{trace},
 \ref{constraints}) becomes:
 
 \bea
 1+ B_1^{(dil)}(x^+, \rho)(\partial_- \rho) &+&   \nn \\
 B_2^{(dil)}(x^+, \rho)(\partial_- \rho) (\partial_+\rho) +
 B_3^{(dil)}(x^+, \rho)(\partial_- \partial_+ \rho) &=& 0, \nn \\
 B_0^{+-}(x^+, \rho) + B_1^{+-}(x^+, \rho)(\partial_- \rho) &+& \nn \\
 B_2^{+-}(x^+, \rho)(\partial_- \rho)(\partial_+ \rho) +
 B_3^{+-}(x^+, \rho)(\partial_- \partial_+ \rho) &=& 0, \nn \\
 B_2^{--}(x^+, \rho)(\partial_- \rho)^2 + B_3^{--}(x^+, \rho)
 (\partial_-^2 \rho) &=& 0, \nn \\
 B_0^{++}(x^+, \rho) + B_1^{++}(x^+, \rho)(\partial_+ \rho) &+& \nn \\
 B_2^{++}(x^+, \rho)(\partial_+ \rho)^2 +
 B_3^{++}(x^+, \rho)(\partial_+^2 \rho) &=& 0,
 \label{beta}
 \eea
 where $B_i^j$'s are expressions of the fields
 $f_{st}(\rho)$ , $A_{st}(\rho)$, $H(x^+,\rho)$, $A(x^+,\rho)$ and
 their derivatives very complicated to be presented here. Following 
 a procedure similar to the one adopted for the static case we eliminate
 the derivatives of the field $\rho$ using  the ansatz
 in (\ref{timeh1}) and the form of the third of the
 equations in (\ref{beta}).
 \newline
 The resultant system is a highly non-linear partial differential system 
 hard to be solved. Instead having in mind the method which has been
 exposed in the case of the static problem we seek solutions in 
 $\kappa$-series expansion. The general solution of this system for the
 fields up to order $\kappa^2$ comes to be the following

 \be
 \phi(x^+,\rho)  = \rho(x^+,x^-) + \frac{\kappa c_0}{4 F_0^2(x^+)},
 \label{timedilaton}
 \ee
and
 \be
 e^{-2\rho}=c_8 e^{-\kappa c_0/2F_0^2(x^+)} + F_0^2(x^+)(c_7-x^-)
 e^{-\kappa c_0/2F_0^2(x^+)} + (8c_4\kappa + 4\kappa^2c_6)F_0^2(x^+),
 \label{timerho}
\ee
where $c_i$'s are integration constants. 
In the above  the integration constant $c_8$ has to be non-negative,
 since when $\kappa=0$ coincides with the mass parameter of the
 Schwarzschild solution. Furthermore the function
 $F_0(x^+)$ satisfies the equation
 
 \be
 F_0^{'}(x^+) = \frac{1}{2F_0(x^+)}e^{\kappa c_0/2F_0^2(x^+)}
 \ee
with solution given in implicit form by

 \be
 x^+ = F_0^2(x^+)e^{-\kappa c_0/2F_0^2(x^+)}-
       \frac{\kappa c_0}{2} E_i(-\frac{\kappa c_0}{2 F_0(x^+)^2}),
 \label{implicit}
 \ee
 where $E_i$ is the exponential integral function.
 from the desired
 asymptotic behaviour of $F_0$,
 $F_0(x^+) \rightarrow 0$ as $x^+ \rightarrow 0$ and
 $F_0(x^+) \rightarrow \sqrt{x^+}$ as  $x^+ \rightarrow +\infty$,
 the constant $c_0$ has to be positive and thus $F_0$ is a
 monotonic function of $x^+$.

 For the Ricci scalar $R=8e^{-2\rho}\partial_+\partial_-\rho$ we get
 the expression

 \be
 R=4e^{2\rho(x^+,x^-)} \{ c_8 e^{\kappa c_0/2F_0^2(x^+)} - 
 2c_0(2c_4\kappa^2+c_6\kappa^3) \}
 \ee

 At this point we notice the close similarity of the above solution
 with the exact solution found in \cite{exact} where the role  of
 the classical tachyon configuration is played here by the
 function $[F_0(x^+)]^{-1}$. This solution
 describes a black hole geometry. The event
 horizon is at some $x^-=constant$.  As far as the physical singularity
 is concerned we see that besides the singularity coming from the
 zero of $e^{-2\rho}$ in (\ref{timerho}) we have the
 curvature blowing up also 
 at  $x^+=0$. As it is explained in \cite{exact} this is not
 a naked singularity but rather an initial one. 
  This extra singularity
 is a general feature of the solution and is inherently
 connected to the deviation from staticity. In (\ref{timedilaton})
 the singularity appears in $\kappa$-order since we introduce
 time-dependence to this order. This could be implemented
 at higher orders in the $\kappa$-expansion and in this case
 we can see that the equation (\ref{implicit}) for the
 corresponding $F_0$ remains the same. If for example
 we change the relation (\ref{timedilaton}) keeping
 the static contribution up to $\kappa^2$ order
 then one can see that exactly the same
 time dependent term $\frac{1}{F_0^2}$, with $F_0$ given
 by (\ref{implicit}), emerges at $\kappa^3$ order, while
 all the static contribution to the dilaton field disappears.
 In fact this is the only kind of time dependence permitted
 if we insist on the analyticity in the parameter $\kappa$.
 Our analysis cannot exclude solutions non-analytic in
 $\kappa$ but such solutions cannot describe an evolution
 scheme between two static black hole geometries since
 these, are analytic in $\kappa$.

 \section{Conclusions}
 In this work we consider solutions of two-dimensional dilaton
 gravity with $R^2$ term which can be viewed as backreaction term
 from quantization of matter. Of course this term has interest
 by itself considered as $R^2$ gravity but we emphasize its
 quantum origin and the fact that, in the semiclassical approximation,
 it does not contribute to the Hawking radiation \cite{hawking}.
 We confirm the static solutions found in \cite{backre} working
 in the conformal gauge and using light-cone coordinates. As a
 bonus of the new analysis we find that the black hole
 covers a part of the physical space of the Schwarzschild
 geometry. In particular the horizon (and eventually the
 sigularity also) lies in this region. Futhermore we
 find that the expansion in powers of the coefficient
 of the $R^2$ term describes well the qualitative
 features of the new geometry. We address also the time dependent
 problem. We seek for solutions analytic
 in $\kappa$ since such solutions have the possibility
 to describe an ordinary evolution scheme between the
 Schwarzschild black hole and the backreacted static one.
 Nevertheless we find that allowing time dependence, even in the
 manner described above,  we have
 a drastic change of the geometry. A new singularity appears
 at $x^+=0$. We note that the solution
 found has black hole structure and this reflects the
  fact that the term included in the action does not contribute
  to the thermal radiation. On the other hand the calculation
  of the
  Bogoliubov coefficient in this background \cite{gang6}, shows
  that we have a non-thermal spectrum and hence a
  thermodynamic instability of the system. We remark here that
  the horizon of the static $R^2$  black hole,
  being removed in the physical spacetime
  region of the Schwarzschild black hole, can in principle
  cover this extra singularity. This in connection
  with the evolution caused from the thermodynamic
  instability shows that the backreaction
  effects are very important for the black hole evolution.
  We conjecture that these effects may prevent the full
  evaporation of the black hole \cite{fabri}. Nevertheless the study
  of such a quantum evolution or any other possibility
  like the one described in \cite{gang6} demands a more
  general context beyond the semiclassical approximation
  adopted here.

 \end{document}